# Microscopic and macroscopic compaction of cohesive powders

M. Morgeneyer, J. Schwedes
*Technical University of Braunschweig, Institute for Mechanical Process Engineering, Braunschweig, Germany*

K. Johnson, D. Kadau, D.E. Wolf
*University Duisburg-Essen, Campus Duisburg, Department of Physics, Duisburg, Germany*

L. Heim
*Max-Planck Institute for Polymer Research, Mainz, Germany*

ABSTRACT: A novel method to investigate the compaction behaviour of cohesive powders is presented. As a sample, a highly porous agglomerate formed by random ballistic deposition (RBD) of micron sized spherical particles is used. A nanomanipulator deforms this small structure under scanning electron microscope observation, allowing for the tracking of individual particle motion. Defined forces are applied and the resulting deformations are measured. The hereby obtained results are compared to results from threedimensional discrete element simulations as well as macroscopic compaction experiments. Relevant simulation parameters are determined by colloidal probe measurements.

## 1 INTRODUCTION

### 1.1 *Compaction of powders*

When particles form a bulk solid, their interparticular forces determine the bulk's initial porosity. External loads, which may include the particles' own weight, then can lead to a further change of porosity, i.e. compaction. For industrial applications – such as powder handling, agglomeration and sintering – it is crucial to achieve an understanding of the physics of the compaction process. A first step here is to establish for beginning compaction constitutive models relating the bulk's porosity $E$ to the compaction stress $\sigma$ using physically reasonable parameters.

### 1.2 *Simulations and macroscopic experiments*

The investigation of early stage uniaxial compaction of powders (compaction at low stresses leading to bulk texture changes without plastic deformation of the primary particles) in computer simulations and macroscopic experiments has led to the power-law

$$E = E_\infty + \left( \frac{F_C}{x_{50}^2 \sigma} \right)^\alpha$$

(Brendel et al., 2003). This model uses the microscopic parameters cohesion force $F_C$ and mean particle diameter $x_{50}$. The exponent $\alpha$ is close to 0.5 for the spherical particles of carbonyl iron powder (CIP), both in simulations as well as in experiments (Morgeneyer et al., 2004), but approaches 0.1 for non-spherical particles such as limestone or marble (Morgeneyer, 2004). Hence, the exponent $\alpha$ seems to be connected to particle geometry. The interparticular cohesion force $F_C$ was found to be close to 0.5 nN for CIP, i.e. about 1/10 of the average of the directly measured values (Heim et al., 2005).

### 1.3 *Microscopic experiments*

Although the link of simulations to macroscopic experiments yields a constitutive model describing the early stage compaction and using microscopic parameters, it turns out that a deeper understanding of the significance of the microscopic parameters $F_C$ and $\alpha$ can only be obtained by investigating the interparticular behaviour as well as observing the compaction process at a microscopic level. Atomic force microscopy (AFM) and scanning electron microscopy (SEM) are state of the art for such observations. The aim of the project described here is to combine these techniques and then to link microscopic and macroscopic compaction experiments and computer simulations.

## 2 METHODS AND SETUP

### 2.1 *Microscopic compaction experiments*

The investigation of the compaction process of microscopic powder samples, i.e. samples composed of about 100 – 1000 particles, makes three steps necessary, each one of them being challenging:
- Generation of a microscopic powder sample (deposit): The deposit has to be of a well-defined reproducible structure allowing for the simulation of similar samples in terms of structure and size.
- Controlled compaction of the deposit: The sample has to be microscopically deformed, observed and key parameters have to be measured.
- The hereby obtained data then have to be analysed. A correlation to simulation data as well as macroscopic experimental data have to be made possible.

#### 2.1.1 *Generation of deposits*

The particles are deposited in a process referred to as random ballistic deposition (Vold, 1959). It is induced by disintegrating a powder sample by the use of a cogwheel in rarefied air (Poppe et al., 1997). The accelerated particles couple to the laminar gas flow which is guided through a filter membrane. The interparticular forces of about 60 nN (Heim et al., 1999) are about five orders of magnitude higher than the mass forces (using 1.5 µm $SiO_2$-particles). The particles' ballistic trajectories end in a hit-and-stick process which leads to agglomerates with high porosities of about $E = 0.85$ (Blum & Schräpler, 2005).

#### 2.1.2 *Nanomanipulator*

The nanomanipulator (Kleindiek, Reutlingen, Germany) allows defined motions of a mounting which is equipped with a cantilever probe for this experiment. This probe is moved by the extension or contraction of a piezo crystal under electrical voltage. The piezo-driven probe allows movements of the tip in 5 nm increments in a rotational movement and 0.5 nm in the manipulator's extension (see Figure 1). This tool allows the precise positioning and movement of an AFM cantilever under SEM observation. Using a low current SEM allows the investigation of pristine samples both of conducting as well as of non-conducting powders. The analysis of the SEM images allows for a direct evaluation of the cantilever's bending and thus for calculating the force applied to the sample.

This makes this set-up an AFM under SEM control.

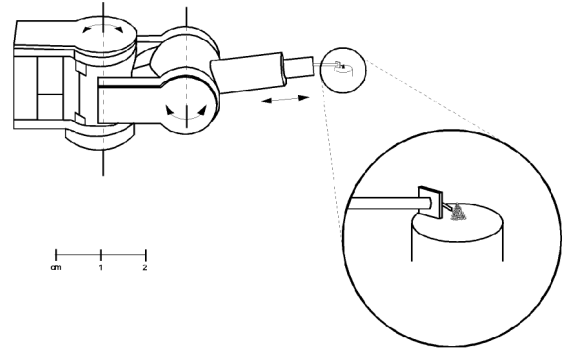

Figure 1. Charting of the nanomanipulator, arrows showing the directions of movement related to the axes of motion

#### 2.1.3 *Data analysis*

The described techniques provide access to information about the positions of individual single particles by continous SEM observation of the sample (Figure 2). As SEM pictures are taken at a sufficiently high frequency, the motion of the individual particles can be traced while the agglomerate is compressed. The forces initiating compression are evaluated. With this information a direct comparison with the simulation is possible.

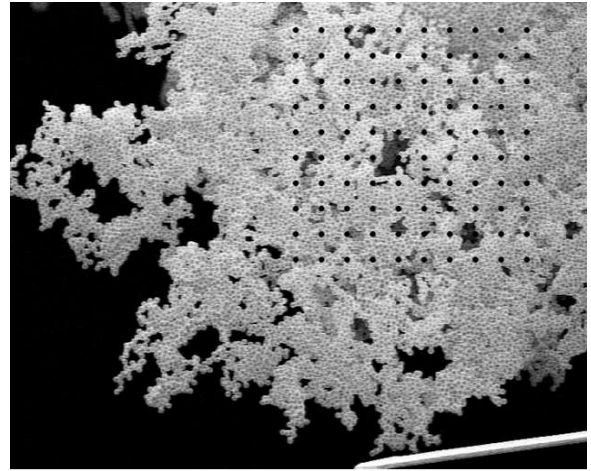

Figure 2. SEM-image of the uncompacted sample, a 10x10 matrix of particles is marked, image width 211 µm

### 2.2 *Simulations*

The simulations are done with a three-dimensional contact dynamics method (Brendel et al., 2004), modelling the particles as perfectly rigid spheres and using the condition of perfect volume exclusion to calculate the force network of the agglomerate. The compaction is strain controlled and the force on the piston is evaluated. For cohesive particles, tension simulations are also accessible.

The low number of particles in the nanoindenter experiment and the access to directly measured interparticle forces by AFM measurements makes it possible to have experiments and simulations with the same number of particles and the same microscopic parameters. Data on particle positions during com-paction are of course available for a direct comparison to the experimental data.

Like in the microscopic experiment, the starting configuration is generated by random ballistic deposition of spherical particles within a cylindrical boundary (Figure 3). A capture radius of one particle diameter models the hit-and-stick process of the experimental deposition and results in the same porosities of approximately $E = 0.85$.

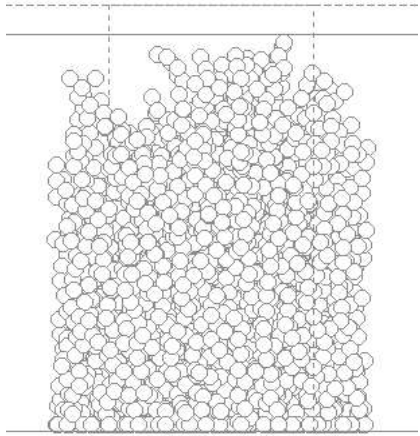

Figure 3. Snapshot of the simulation at the beginning of compaction. A piston is moved down vertically onto a cylindrical free-standing RBD (no side boundaries).

## 2.3 *Macroscopic experiments*

Macroscopic compaction experiments are carried out using the true biaxial shear tester (Harder, 1986), which allows for compacting powder samples of volumes of about ½ litre, the samples remaining brick shaped during the whole compaction procedure (Figure 4). The side, bottom and top plates are covered with lubricated highly flexible rubber membranes in order to avoid friction stresses. Thus, the axis of principle stresses and principle strains coincide. For compaction, two opposing side plates are moved towards each other. Both stresses (up to 30 kPa) and strains (up to 40%) are continously measured.

For the powder compaction's investigation at higher stresses (up to 300 kPa), a uniaxial compaction apparatus (Morgeneyer et al., 2004) is used.

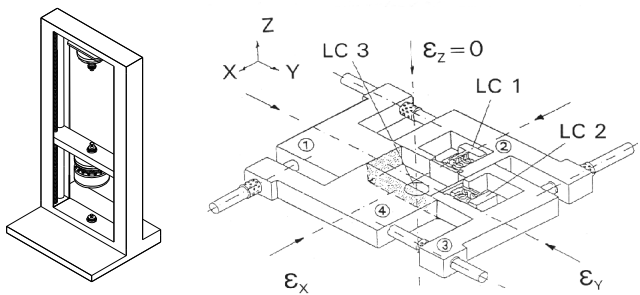

Figure 4. The uniaxial compaction apparatus (sample size 40x100 mm, left) and the true biaxial shear tester (sample size 130x70x35 mm, right)

## 3 RESULTS

### 3.1 *Microscopic results*

The force on the cantilever increases approximately linearly during compaction (Figure 6). In principle the cantilever position can be translated to a porosity, because the motions of the individual particles have been traced. There are distinct events of sudden force decrease which indicate the existence of substructures with higher stability.

When the cantilever is retracted, the cohesive powder exerts a tensile force on it.

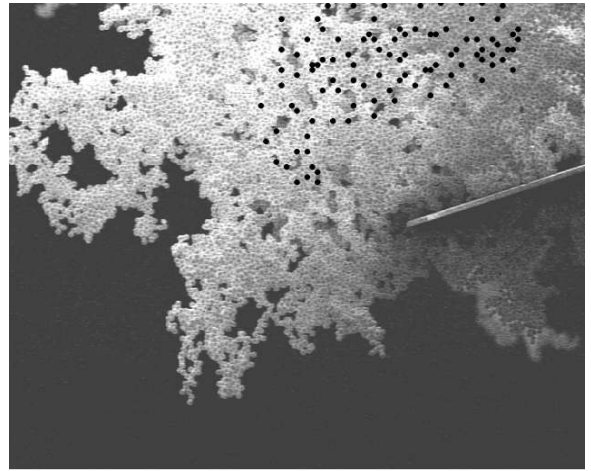

Figure 5. SEM-image of the compacted sample, deformed 10x10 matrix (cf. Figure 2), image width 211 μm

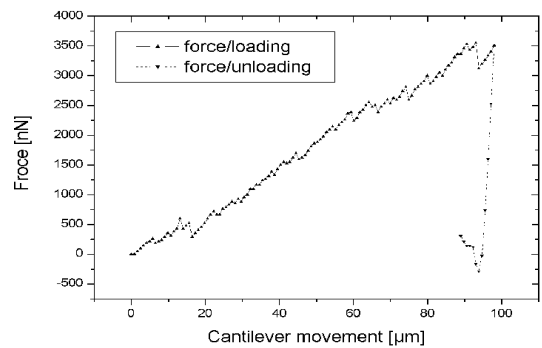

Figure 6. Force-displacement curve of microscopically compacted $SiO_2$ particles

### 3.2 *Simulation results*

Initial simulations were done with random ballistic deposits of approximately 2000 particles (Figure 7). After a transient phase during which only surface branches of the RBD are compacted, the piston starts compacting the bulk. Then qualitatively the same linear force increase as in the nanoindenter experiments is obtained, but with strong force fluctuations coinciding with the forming and breaking of inter-particle contacts. The inter-particle cohesion force $F_C$ was set higher in the simulation than in the experiment in order to amplify this effect. The highest spikes lie in the same order of magnitude as $F_C$.

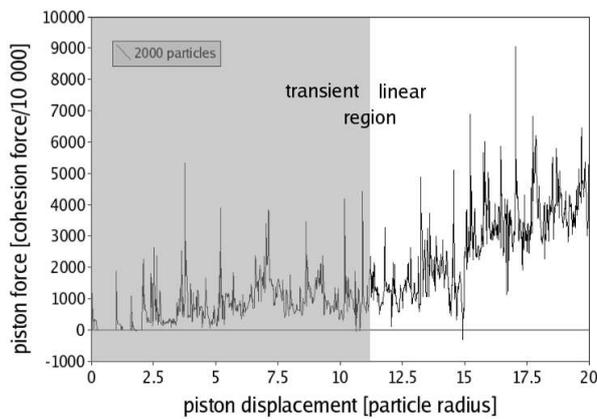

Figure 7. The graph shows the piston force as a function of the piston displacement (simulation)

### 3.3 *Comparison to macroscopic experimental results*

Macroscopically, forces of about 10 kN can be applied to the sample. In Figure 8 only data up to a force corresponding to a compaction stress of about 13 kN/m² are presented, whereas the maximum stress that can be applied to the microscopic sample can only reach about 1 kN/m$^2$. Nevertheless, the macroscopic experiment reveals for beginning compaction a linear increase of the compaction force, similar to the one observed both in the microscopic experiment and in the simulation. Exponential force increase at further compaction is due to the increasing bulk density.

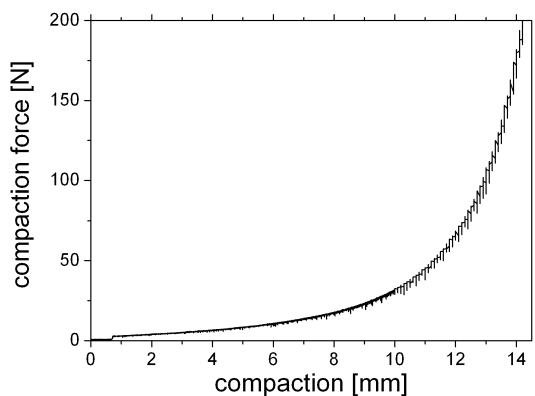

Figure 8. Macroscopic compaction of CIP, uniaxial compaction apparatus, initial height 40 mm

## 4 CONCLUSION AND OUTLOOK

It has been demonstrated that the combination of microscopic and macroscopic experiments and simulations has high potential of providing new insights into the compaction of cohesive granular matter. The possibility to trace individual particles experimentally provides a direct way of validating the simulation models. For this purpose we are working on the generation of deposits with a smaller amount of particles and a well-defined geometry. First results are promising.

In order to investigate the influence of microscopic parameters on the compaction process, different powder materials will be investigated both by the nanoindenter method as well as macroscopically by the true biaxial shear tester. The task of a validated and well calibrated simulation model will be to assist the microscopic experiments e.g. with specifying the boundary conditions and the stress state.

Besides compaction of powder samples, tensile tests can also be realized, see (Morgeneyer & Schwedes, 2004). This option exists also for the microscopic experiment, as shown by the unloading data in Figure 6.

## 6 ACKNOWLEDGEMENTS

We gratefully acknowledge J. Blum and R. Schräpler; Institute for Geophysics and Extraterrestrial Physics, TU Braunschweig, Germany, for the preparation of the deposits as well as H.-J. Butt and M. Kappl, Max-Planck-Institute for Polymer Research, Polymer Physics, Mainz, for their support.

Financial support from German Research Foundation (DFG) within "Verhalten granularer Medien" is gratefully acknowledged.